\documentclass[twocolumn,preprintnumbers,aps,pra,superscriptaddress,amsmath,amssymb]{revtex4-1}

\usepackage{graphicx}
\usepackage{dcolumn}
\usepackage{bm}
\usepackage{bbm}
\usepackage{hyperref}
\usepackage{color}
\usepackage{multirow}
\usepackage{physics}

\usepackage{natbib}

\newcommand{\Ham}{\hat{H}}
\newcommand{\Prop}{\hat{U}}

\newcommand{\sigx}{\hat{\sigma}_x}

\newcommand{\sigz}{\hat{\sigma}_z}
\newcommand{\sigi}{\hat{\sigma}_i}

\newcommand{\Sz}{\hat{S}_z}

\usepackage{braket}

\begin{document}

\title{The non-vanishing effect of detuning errors in dynamical decoupling based quantum sensing experiments} 

\author{J. E. Lang}
\email{jacob.lang.14@ucl.ac.uk}
\affiliation{Department of Physics and Astronomy, University College London, Gower Street, London WC1E 6BT, United Kingdom}

\author{T. Madhavan}
\affiliation{School of Physics, The University of Melbourne, VIC 3010, Australia}

\author{J.-P. Tetienne} 
\email{jtetienne@unimelb.edu.au}
\affiliation{School of Physics, The University of Melbourne, VIC 3010, Australia}	

\author{D. A. Broadway}
\affiliation{School of Physics, The University of Melbourne, VIC 3010, Australia}
\affiliation{Centre for Quantum Computation and Communication Technology, School of Physics, The University of Melbourne, VIC 3010, Australia}

\author{L. T. Hall}
\affiliation{School of Physics, The University of Melbourne, VIC 3010, Australia}

\author{T. Teraji}
\affiliation{National Institute for Materials Science, Tsukuba, Ibaraki 305-0044, Japan}


\author{T. S. Monteiro}
\affiliation{Department of Physics and Astronomy, University College London, Gower Street, London WC1E 6BT, United Kingdom}

\author{A. Stacey}
\affiliation{School of Physics, The University of Melbourne, VIC 3010, Australia}
\affiliation{Centre for Quantum Computation and Communication Technology, School of Physics, The University of Melbourne, VIC 3010, Australia}

\author{L. C. L. Hollenberg}
\affiliation{School of Physics, The University of Melbourne, VIC 3010, Australia}
\affiliation{Centre for Quantum Computation and Communication Technology, School of Physics, The University of Melbourne, VIC 3010, Australia}

\date{\today}
	
\begin{abstract}
Characteristic dips appear in the coherence traces of a probe qubit when dynamical decoupling (DD) is applied in synchrony with the precession of target nuclear spins, forming the basis for nanoscale nuclear magnetic resonance (NMR). The frequency of the microwave control pulses is chosen to match the qubit transition but this can be detuned from resonance by experimental errors, hyperfine coupling intrinsic to the qubit, or inhomogeneous broadening. The detuning acts as an additional static field which is generally assumed to be completely removed in Hahn echo and DD experiments. Here we demonstrate that this is not the case in the presence of finite pulse-durations, where a detuning can drastically alter the coherence response of the probe qubit, with important implications for sensing applications. Using the electronic spin associated with a nitrogen-vacancy centre in diamond as a test qubit system, we analytically and experimentally study the qubit coherence response under CPMG and XY8 dynamical decoupling control schemes in the presence of finite pulse-durations and static detunings. Most striking is the splitting of the NMR resonance under CPMG, whereas under XY8 the amplitude of the NMR signal is modulated. Our work shows that the detuning error must not be neglected when extracting data from quantum sensor coherence traces. 

\end{abstract}

\maketitle

\section{Introduction}

In the last decade, the nitrogen-vacancy (NV) centre in diamond~\cite{Doherty2013} has emerged as a leading qubit system for the development of quantum technologies. The optical addressability and long coherence times (even at room temperature) of its electronic spin make the NV an excellent platform for quantum sensing \cite{rondin2014,schirhagl2014nitrogen,suter2016single,degen2017quantum}, computing \cite{childress2006coherent, ladd2010quantum, taminiau2014universal} and devices \cite{breeze2018continuous}. A particularly promising application is nanoscale nuclear magnetic resonance (NMR), which relies on the ability of the NV to detect the weak oscillating signals from target nuclear spins using dynamical decoupling (DD) schemes~\cite{DeLange2010,Meriles2010,suter2016single,degen2017quantum,Perunicic2016}. These methods have been used to detect single nuclear spins and spin clusters inside the diamond \cite{zhao2012sensing, kolkowitz2012sensing, taminiau2012detection,shi2014sensing}, ensembles of nuclear spins on the diamond surface \cite{staudacher2013nuclear,mamin2013nanoscale,Loretz2014,DeVience2015} and ultimately single nuclear spins on the diamond surface \cite{muller2014nuclear, sushkov2014magnetic, lovchinsky2016nuclear}. DD is also utilised in protocols for increasing spectral resolution \cite{laraoui2013high, Staudacher2015,kong2015towards,boss2017quantum,schmitt2017submillihertz,Glenn2018} and controlling nuclear spins in spin registers \cite{childress2006coherent, taminiau2014universal}. 

NMR detection with DD relies on the fact that DD normally protects a qubit state from decoherence by averaging out the effects of environmental noise  \cite{hahn1950spin, carr1954effects, meiboom1958modified}. When the DD pulses are applied in synchrony with a nuclear spin signal, the decoupling fails and characteristic dips appear in coherence traces. The position and depth of these dips are used to extract information about the incident signal, such as the number and spatial location of the target spins~\cite{muller2014nuclear,lovchinsky2016nuclear,pham2016nmr}. It is thus vitally important to be able to accurately model the sensor coherence response, including in the presence of unavoidable control errors. 

In this work, we focus on the effect of a specific type of error in DD-based NMR sensing, namely detuning errors. A static detuning is present whenever the microwave frequency within each DD pulse does not match the qubit transition frequency, which for the NV corresponds to one of the $m_s = 0 \leftrightarrow m_s = \pm 1$ spin transitions. Detunings emerge from experimental errors in the microwave driving frequency but also from more intrinsic sources that shift the qubit frequency itself, such as hyperfine splittings from the host nitrogen spin of the NV as well as inhomogeneous broadening due to the fluctuating environment. For instantaneous pulses, detuning errors can generally be neglected when modelling Hahn echo and DD experiments as the static field is completely refocussed after each pulse. This is how the Hahn echo and DD protocols extend the coherence of the NV spin from the dephasing time, $T_2^*$, to the true coherence time, $T_2$~\cite{DeLange2010}. Here we show, however, that in the presence of finite pulse-durations~\cite{Loretz2015}, the microwave detuning error can no longer be ignored and can have drastic effects on the sensor coherence response. In particular, we find that detunings can split the NMR resonance in CPMG based detection, and modulate the resonance amplitude under the XY8 sequence. We investigate these effects both analytically, by deriving new expressions for the sensor coherence response to a classical field under finite-duration-pulse control, and experimentally, by detecting an ensemble of protons spins prepared on the diamond surface using near-surface NV centres. Our findings have immediate ramifications for experiments that rely on the position or strength of the coherence dip to extract information from the environment. More broadly, these effects present an interesting class of problems as they cannot be captured by instantaneous pulse models. Whilst many studies have focussed on the decoupling efficiency of different DD sequences to errors \cite{maudsley1986modified, gullion1990new,Shim2012,Wang2012,Ali2013,Farfurnik2015, casanova2015robust}, our work represents a new study of the detuning effect on the resonant dip associated with nuclear spin detection.  

The manuscript is organised as follows. We first detail our methods (Sec.~\ref{sec:methods}) for the theoretical description of DD sequences in the presence of static detunings and finite-duration pulses (\ref{sec:meth theory}) and for the experimental verification (\ref{sec:meth exp}). In Sec.~\ref{sec: CPMG}, we present the results in the case of the CPMG sequence, whereas Sec.~\ref{sec: XY8} is devoted to the XY8 sequence. Finally, we discuss the implications of these findings and future work (Sec.~\ref{sec:conclusion}).   

\begin{figure*}[t!]
\begin{center}
\includegraphics[width=0.9\textwidth]{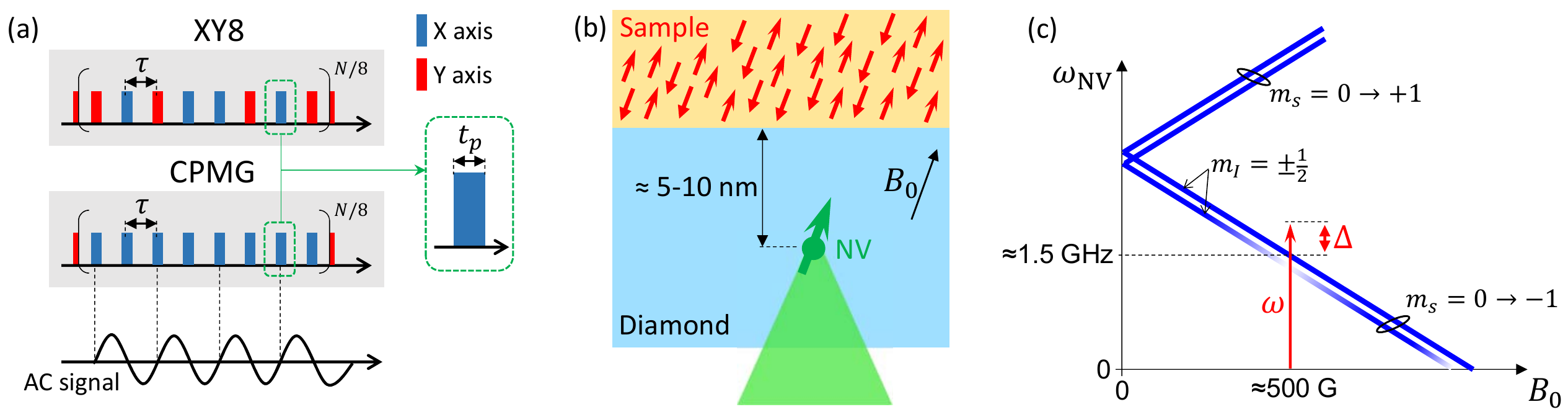}
\caption{(a) The two dynamical decoupling pulse sequences considered in this work to detect an oscillating (AC) magnetic signal generated, for instance, by an ensemble of nuclear spins. The blue and red pulses correspond to $0^\circ$ and $90^\circ$ phase shift of the microwave driving field, respectively. (b) Schematic of the experiment: a single nitrogen-vacancy (NV) centre below the diamond surface, with a semi-infinite sample of nuclear spins above the surface. (c) Transition frequencies of the NV electronic spin as a function of the axial magnetic field $B_0$. The hyperfine structure due to the $^{15}$N nuclear spin of the NV centre ($m_I=\pm\frac{1}{2}$) is shown, with the shading of the lines denoting the amount of nuclear spin polarisation under optical pumping, in particular near the ESLAC at $B\approx500$G the nuclear spin is almost fully polarised into the $m_I=-\frac{1}{2}$ state~\cite{Jacques2009}.}
\label{FigIntro}
\end{center}
\end{figure*}

\section{Methods} \label{sec:methods}

\subsection{Theory} \label{sec:meth theory}

The detection of a large ensemble of weakly coupled nuclear spins is well described by a semi-classical model, which treats the sensor quantum mechanically but represents the nuclear spin signal as a classical field~\cite{pham2016nmr, zhao2012sensing}. To include the effect of detuning errors and finite-duration pulses, we derive here a new expression for the sensor coherence response. We consider a spin qubit subject to a classical signal and dynamical decoupling control (Fig.~\ref{FigIntro}a). The classical signal can be modelled by a time-dependent magnetic field, $\textbf{B}(t) = (B_x(t), B_y(t), B_z(t))$, but in the presence of a large external magnetic field applied along the $z$-axis -- or, in the case of the NV centre, due to a large zero-field splitting -- only the $z$-component of the signal field survives the pure-dephasing approximation. The dynamical decoupling is modelled by the pulse Hamiltonian, $\Ham_p(t) = \Delta\Sz + \sum_{m = 1}^N \Omega(t - t_m)\hat{S}_{\phi_m}$, which describes a sequence of $N$ microwave pulses at times $t_m$ (spaced by $\tau$) about the axes described by the phase $\phi_m$ and with a shape $\Omega(t)$ which is zero outside some width $t_p$. $\Delta$ is the microwave detuning from resonance (see Fig.~\ref{FigIntro}c) and the pulse Hamiltonian is presented in the frame rotating with the microwave frequency (neglecting counter-rotating terms).

The Hamiltonian can be written in the toggling frame, the frame rotating under $\Ham_p(t)$ (see Appendix~\ref{app: classical}), as
\begin{equation}
\Ham(t) = -\frac{\gamma_\text{e}}{2}B_z(t)\sum_i f_i(t) \sigi,
\label{eq: togg frame Ham}
\end{equation}
where $\gamma_\text{e} = -28~\text{GHz/T}\times 2\pi$ is the gyromagnetic ratio of the qubit and $\sigi$ are the usual Pauli matrices. The $f_i(t)$ are the modulation functions which generalise the single, square-wave modulation function used in many semi-classical and quantum models \cite{alvarez2011measuring, zhao2012sensing, cywinski2008enhance, ma2016angstrom, albrecht2015filter, casanova2015robust}. Due to the finite duration, $t_p$, of the microwave $\pi$-pulses, the qubit state is not instantaneously inverted but has some finite-duration transit. In the limit of $t_p \rightarrow 0$, the \textit{parallel modulation function}, $f_z(t)$, recovers this stepped modulation function whilst the \textit{perpendicular modulation functions}, $f_{x,y}(t)$, vanish. For finite pulse-durations, however, the perpendicular modulations are non-zero introducing some spin-mixing into a previously pure-dephasing Hamiltonian. These generalised modulation functions have been  described previously \cite{lang2017enhanced} but here we include a static microwave detuning in the pulse Hamiltonian which modifies their behaviour. 

To model the coherence response under this finite pulse control, we follow closely the derivation for the instantaneous pulse case \cite{pham2016nmr,cywinski2008enhance,zhao2012sensing}. We model a typical DD experiment that begins with the sensor in an initial superposition state $\ket{\psi(0)} = (\ket{u} + \ket{d})/\sqrt{2}$ (where $\ket{u,d}$ are the \textit{up} and \textit{down} states of the sensor qubit). After the application of a DD sequence of total length $t_\text{total}$ the coherence is measured along the original superposition axis, $\mathcal{L}(t_\text{total}) = \braket{\psi(t_\text{total})|\sigx|\psi(t_\text{total})}$. The full evolution is determined by $\ket{\psi(t_\text{total})} = \Prop(t_\text{total})\Prop_p(t_\text{total})\ket{\psi(0)}$ where $\Prop(t)$ is the propagator (given in Appendix~\ref{app: classical}) associated with the toggling frame Hamiltonian, Eq.~\eqref{eq: togg frame Ham} and $\Prop_p(t)$ is the pulse propagator associated with the pulse Hamiltonian $\Ham_p(t)$.

To observe detuning error effects on the resonant coherence dip without the loss of background coherence we assume that the error is small enough to still satisfy $\Prop_p(t_\text{total})\ket{\psi(0)} \simeq \ket{\psi(0)}$. For the XY8 sequence this is satisfied by its robust design which applies pulses at different phases to cancel the accumulation of errors - up to second order \cite{wang2012comparison}. For the CPMG sequence the error accumulates quickly but results in an effective rotation about the $x$-axis \cite{wang2012comparison}. However, by choosing the initial state as $\ket{\psi(0)} = (\ket{u} + \ket{d})/\sqrt{2}$ this rotation has no effect and $\Prop_p(t_\text{total})\ket{\psi(0)} \simeq \ket{\psi(0)}$ is still valid. When the initial state is not aligned with the $x$-axis the approximation fails.

We find that, after a DD sequence of total length $t_\text{total}$, the coherence is 
\begin{equation}
\mathcal{L}(t_\text{total}) = 
\exp(-\frac{1}{2}\frac{\gamma_\text{e}^2}{2\pi}\int_{-\infty}^\infty S(\omega)|\tilde{\textbf{f}}(\omega)|^2 d\omega \ t_\text{total}^2),
\label{eq: L classical}
\end{equation}
where $S(\omega)$ is the noise spectrum of the classical signal (resulting from the magnetic field
$B_z(t)$) and $\tilde{\textbf{f}}(\omega) = (\tilde{f}_x(\omega), \tilde{f}_y(\omega), \tilde{f}_z(\omega))$ where $\tilde{f}_i(\omega) = \frac{1}{t_\text{total}}\int_0^{t_\text{total}}f_i(t)\exp(-i\omega t)dt$ is the spectrum of the $i$-th modulation function. This expression can be interpreted the standard way -- the DD sequence creates a narrow-band filter (with filter function $|\tilde{\textbf{f}}(\omega)|^2$) that is scanned across the bath noise spectrum as the pulse spacing $\tau$ is increased. For instantaneous pulses, Eq.~\eqref{eq: L classical} reduces exactly to the usual semi-classical expression as $\tilde{f}_{x,y}(\omega) = 0$ and the filter function is simply $|\tilde{\textbf{f}}(\omega)|^2=|\tilde{f}_z(\omega)|^2$~\cite{zhao2012sensing}. The inclusion of detuning errors and finite-duration pulses alters this filter function and thus the coherence response. In Sec.~\ref{sec: CPMG}, we specifically analyse the response of the CPMG sequence to detuning errors in the presence of finite-duration pulses. We study the modulation functions and their spectra to predict the effect of the coherence response -- a drastic splitting of the characteristic dip. This prediction is verified experimentally. In Sec.~\ref{sec: XY8}, we examine the effect of detuning errors on the coherence signals under XY8 control. Both instances have implications for experiments that extract information from the dip position or strength. 

\subsection{Experiment} \label{sec:meth exp}

\begin{figure*}[t!]
\begin{center}
\includegraphics[width=0.8\textwidth]{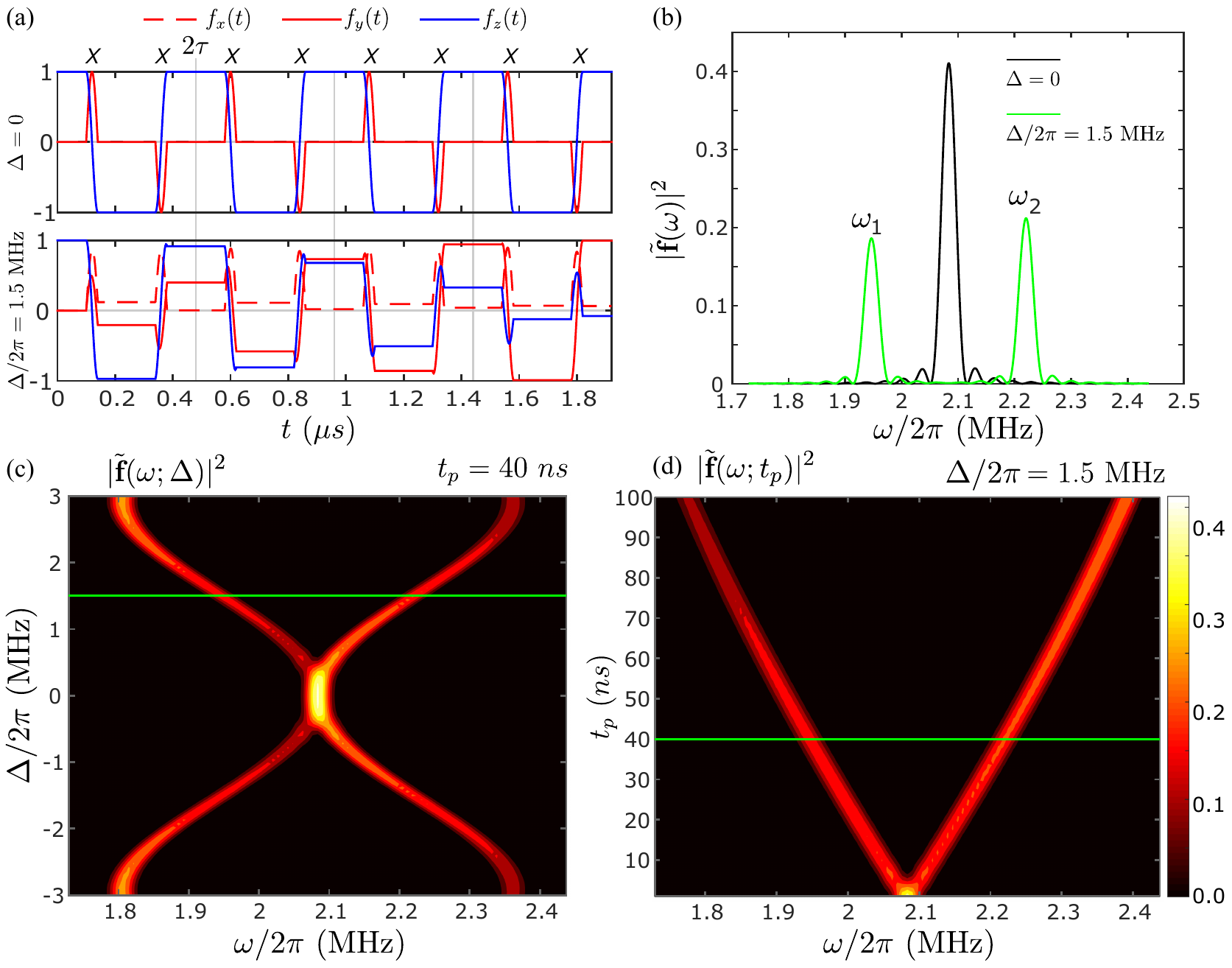}
\caption{
(a) Modulation functions $(f_x,f_y,f_z)$ in the time domain for the CPMG sequence with finite duration pulses for zero and non-zero detuning error. The simulation here has pulse spacing $\tau = 240$ ns, pulse width $t_p = 40$ ns and detunings $\Delta = 0$ (top graph) and $\Delta/2\pi = 1.5$ MHz (bottom). These values were chosen to be representative of the experiment performed in Fig.~\ref{FigCPMG}. (b) Filter function, $|\tilde{\textbf{f}}(\omega)|^2$, for the CPMG sequences shown in (a) but with a total of $N = 128$ pulses. For zero detuning there is a central peak at $\omega_\text{DD}/2\pi = 2.08$ MHz which splits at non-zero detuning. (c) Scan of the filter function shown in (b) as a function of the detuning strength, $\Delta$. (d) Scan of the filter function shown in (b) as a function of the pulse width, $t_p$. The green solid lines indicate the equivalent slices in each map and these also correspond to the filter function shown in (b).
}
\label{fig: CPMG}
\end{center}
\end{figure*}

To experimentally test our predictions, we performed measurements on single NV centres implanted in a (001)-oriented electronic-grade diamond purchased from Delaware Diamond Knives and overgrown with $2~\mu$m of $^{12}$C-enriched (99.95\%) diamond~\cite{Teraji2015}. $^{15}$N$^+$ ions were implanted (InnovIon) at a fluence of $10^9$~ions/cm$^2$ and energy of 3 keV, corresponding to NV depths in the range $5-10$~nm~\cite{Wood2017,Broadway2018b}. Following implantation, the diamond was annealed at 950$^\circ$C for 2h in a vacuum of $\sim10^{-5}$~Torr, acid cleaned (15 minutes in a boiling mixture of sulphuric acid and sodium nitrate) and annealed at 500$^\circ$C for 4h in an oxygen atmosphere~\cite{lovchinsky2016nuclear,Tetienne2018}.

As a target sample, we applied a layer of immersion oil to the diamond surface (Fig.~\ref{FigIntro}b), resulting in an effectively semi-infinite bath of proton spins with a density of about 60~nm$^{-3}$~\cite{staudacher2013nuclear,pham2016nmr}. An external magnetic field ${\bf B}_0$ was applied along the NV axis to lift the degeneracy between the $m_s=\pm1$ spin sublevels, with the microwave driving field (frequency $\omega$) approximately resonant with the $m_s = 0 \leftrightarrow-1$ transition (Fig.~\ref{FigIntro}c). The magnetic field was applied using a temperature-controlled permanent magnet to minimise magnetic field drifts~\cite{Broadway2018}. Due to hyperfine coupling with the $^{15}$N nuclear spin of the NV (a spin-$\frac{1}{2}$), the electronic spin transition $m_s = 0 \leftrightarrow-1$ has in fact two possible frequencies depending on the state of the nuclear spin, $m_I=\pm\frac{1}{2}$, separated by about 3 MHz~\cite{Jacques2009}. This gives rise to an intrinsic detuning since the driving microwave field cannot be resonant with both transitions simultaneously. In this paper, however, we chose the strength of the external magnetic field to be near the excited state level anti-crossing (ESLAC), i.e. $B_0\approx500$~G, where the $^{15}$N spin is efficiently polarised into the $m_I=-\frac{1}{2}$ state under optical pumping~\cite{Jacques2009} (Fig.~\ref{FigIntro}c). This allows us to study the case of a single qubit frequency, denoted as $\omega_{\rm NV}$, with an independent control over the detuning $\Delta=\omega-\omega_{\rm NV}$. Another motivation for working near the ESLAC is to facilitate the alignment of ${\bf B}_0$, which can be aligned within a few degrees of the NV axis by using the angle-dependent photoluminescence (PL) induced by the ESLAC~\cite{Wood2016,Tetienne2012,Epstein2005}.

The microwave field was applied using a loop antenna placed in proximity of the diamond. The antenna was connected to a signal generator (Rohde \& Schwarz SMBV100A) gated through the built-in IQ modulation by a pulse pattern generator (SpinCore PulseBlasterESR-PRO 500 MHz) to generate the DD sequences. The microwave pulse shape is roughly flat topped with a measured rise/fall time of $<4$~ns, limited by the bandwidth of the IQ modulation. For each NV studied, we first recorded an optically detected magnetic resonance (ODMR) spectrum at low microwave power to determine the NV transition frequency, $\omega_{\rm NV}$. We then performed a Rabi measurement at high microwave power and zero detuning to determine the $\pi$-pulse duration ($t_p$) used in the subsequent DD measurements, setting $t_p$ at the position of the first trough seen in the Rabi oscillations. When a detuning was applied during a DD measurement, the microwave power and pulse duration $t_p$ were kept constant as determined at zero detuning. We note that the Rabi frequency of the pulses, $\Omega=\pi/t_p$, was always larger than the detuning. We apply pulse durations in the range $t_p = 20 \rightarrow 80$~ns so that $\Omega/2\pi = 6.25 \rightarrow 25$~MHz and test detunings in the range $|\Delta|/2\pi= 0 \rightarrow 3$~MHz.

To account for drifts in the NV frequency, we measured $\omega_{\rm NV}$ via ODMR both before and after the DD measurement and defined the detuning as the difference $\Delta=\omega-\bar{\omega}_{\rm NV}$ where $\bar{\omega}_{\rm NV}$ is the mean NV transition frequency. Drifts in $\omega_{\rm NV}$ of up to 0.2~MHz were observed over the course of a few hours (a typical acquisition time for a single NMR spectrum), mainly caused by residual temperature fluctuations of the magnet~\cite{Broadway2018}. As such, an uncertainty of $\pm0.1$~MHz is associated with the quoted values of the detuning $\Delta$. In the DD experiments, we scanned the pulse spacing $\tau$ and measured the difference $\Delta S=(S_0-S_1)$ where $S_0$ ($S_1$) is the photon count after the DD sequence when the final $\pi/2$ pulse projects the NV spin coherence onto the $|0\rangle$ ($|-1\rangle$) state. The spectrum is then normalised to the $\tau=0$ case, giving a measure of the spin coherence. All measurements are carried out at room temperature.

\section{Effects of detuning in CPMG based NMR}
\label{sec: CPMG}

\begin{figure*}[t!]
\begin{center}
\includegraphics[width=0.7\textwidth]{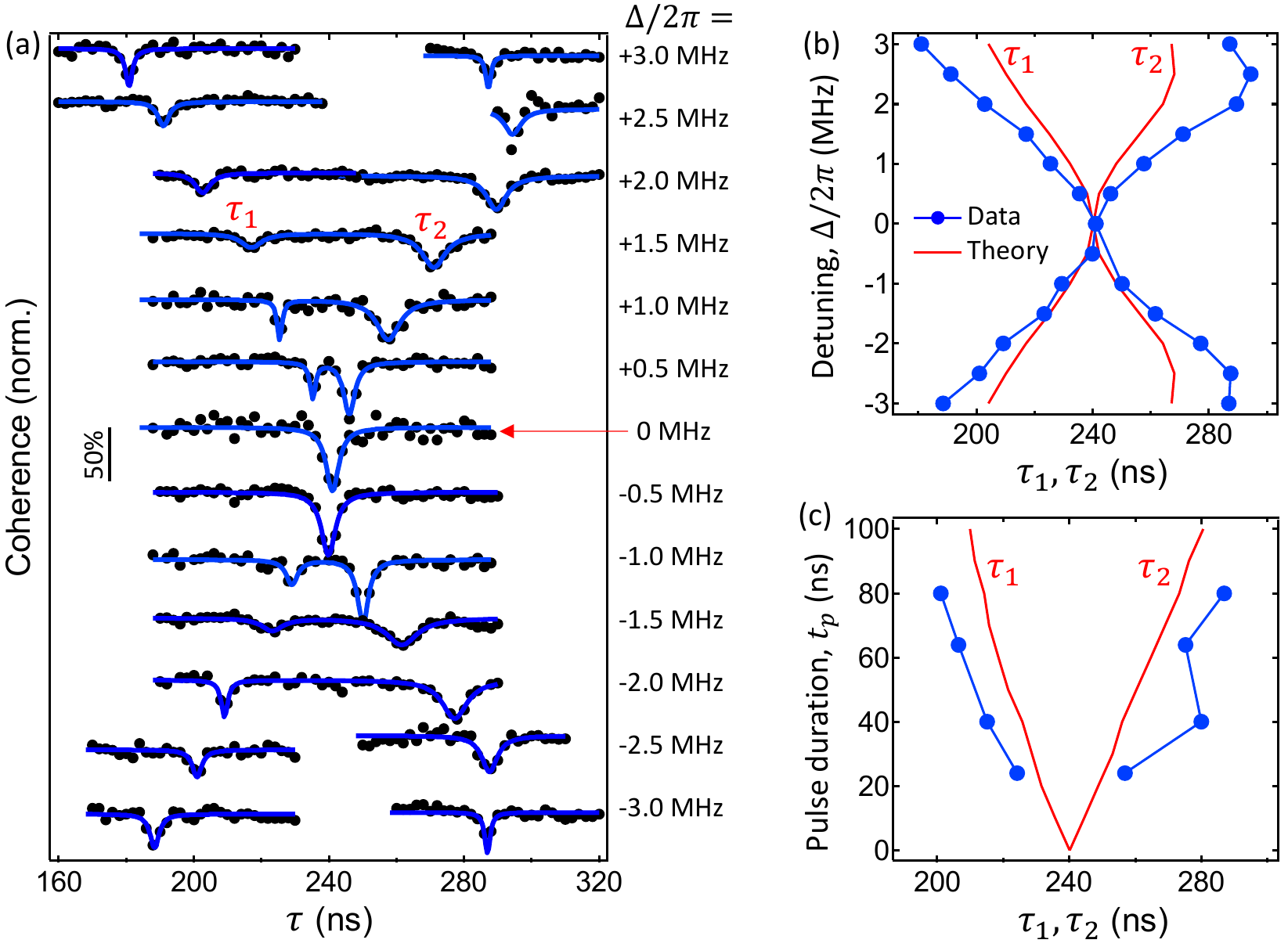}
\caption{(a,b) NMR spectra obtained using the CPMG sequence for various values of the detuning $\Delta$, with a $\pi$-pulse duration $t_p=40$~ns, a number of $\pi$ pulses $N=336$, and under a magnetic field $B\approx500$~G. The different spectra are vertically offset for clarity. The solid lines are double-Lorentzian fits used to estimate the positions of the two NMR dips, $\tau_1$ and $\tau_2$. (b) Positions of the two NMR dips extracted from (a), $\tau_1$ and $\tau_2$, as a function of the detuning $\Delta$. (c) Positions of the two NMR dips as a function of the $\pi$-pulse duration $t_p$ for a fixed detuning $\Delta/2\pi=+1.5$~MHz. Here the number of pulses is $N=256$ (shallower NV compared to (a,b)). In (b,c), the blue dots are the experimental data (blue lines are a guide to the eye) while the red lines are the theory.}
\label{FigCPMG}
\end{center}
\end{figure*}

Figure~\ref{fig: CPMG}a shows the CPMG modulation functions in the time domain for finite-duration pulses with and without a detuning error. For zero detuning, these modulation functions repeat with a 2-pulse period and have the resonant frequency $\omega_\text{DD} = \pi/\tau$ where $\tau$ is the pulse spacing. For non-zero detuning, each pulse accumulates a small error and this imprints a slow oscillation onto the modulation functions (this can be best seen in $f_z(t)$).
Whilst we assumed that the small detuning error had no effect on the coherence background (i.e. $\Prop_p(t_\text{total})\ket{\psi(0)} \simeq \ket{\psi(0)}$) due to it creating an effective $x$-rotation which does not affect the initial sensor state \cite{wang2012comparison}, this rotation \textit{does} cause errors to accumulate in the modulation functions (described in Appendix~\ref{app: classical}).
This beating oscillation splits the resonant frequency of the CPMG sequence. Figure~\ref{fig: CPMG}b shows the filter function $|\tilde{\textbf{f}}(\omega)|^2$ resulting from these modulation functions, which have a single peak at $\omega=\omega_\text{DD}$ for zero detuning and two peaks for non-zero detuning due to the accumulation of the pulse error. These two peaks at $\omega_1$ and $\omega_2$ are well resolved, i.e. the difference $\delta\omega=\omega_2-\omega_1$ is much larger than the peak width (300 kHz against less than 50 kHz, for a detuning $\Delta/2\pi = 1.5$~MHz). They have slightly different amplitudes, which are both roughly half of the amplitude of the non-split peak.

In Fig.~\ref{fig: CPMG}c and \ref{fig: CPMG}d, the dependence of the filter function on detuning strength $\Delta$ and pulse width $t_p$ is presented. The behaviour is symmetric about $\Delta = 0$ but not about the nominal resonant frequency $\omega_\text{DD}$. Indeed, although the two peaks $\omega_1$ and $\omega_2$ shift symmetrically in terms of position, with a non-trivial dependence on $\Delta$, their amplitudes differ, with the $\omega_2$ resonance being stronger than $\omega_1$ for detunings up to $|\Delta/2\pi|\approx2$~MHz but weaker for larger detunings (Fig.~\ref{fig: CPMG}c). The pulse width scan for a constant non-zero detuning (Fig.~\ref{fig: CPMG}d) reveals that the peak positions scale approximately linearly with $t_p$, with again an asymmetry in amplitude. Importantly, we see that instantaneous pulses ($t_p=0$) are unaffected by the detuning (i.e. there is no splitting) and thus cannot be used to accurately model the effect of detuning.

\begin{figure*}[t!]
\begin{center}
\includegraphics[width=0.99\textwidth]{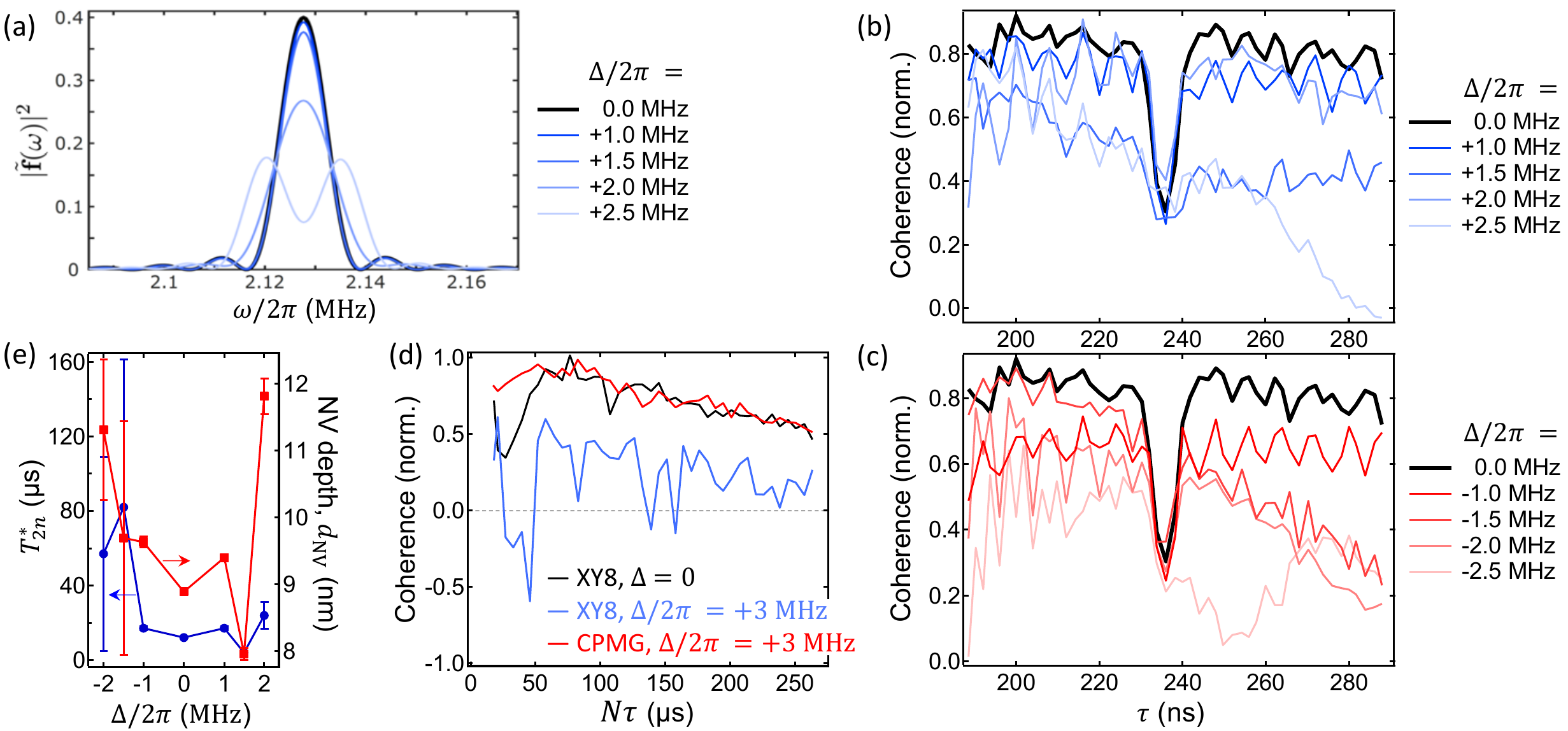}
\caption{(a) Filter function $|\tilde{\bf f}(\omega)|^2$ for the XY8 sequence for various values of the detuning $\Delta$, with $N=376$, $t_p = 40$ ns and $\tau=235$~ns. (b,c) NMR spectra recorded from the same NV as in Fig.~\ref{FigCPMG}a using the XY8 sequence with $t_p=40$~ns and $N=376$, for positive (b) and negative (c) detunings. (d) Decoherence curves recorded for a different NV under the XY8 sequence, with $N=128$, $t_p=40$~ns and a detuning $\Delta=0$ (black line) or $\Delta/2\pi=+3$~MHz (blue). Also shown for comparison is the case of the CPMG sequence with $\Delta/2\pi=+3$~MHz (red). (e) Parameters $d_{\rm NV}$ and $T_{2n}^*$ obtained by fitting the model of Ref.~\cite{pham2016nmr} to the spectra shown in (b,c), for detunings between -2 and +2 MHz. The lines are a guide to the eye. The vertical error bars are the standard error from the fit.}
\label{FigXY8}
\end{center}
\end{figure*}

To test these predictions experimentally, we measured the NV spin coherence while scanning the pulse spacing $\tau$, which is equivalent to scanning the central frequency of the filter function, $\omega_\text{DD} = \pi/\tau$. Since the proton ensemble generates a fluctuating signal peaked at the Larmor frequency $\omega_L/2\pi\approx2.1$~MHz, we expect a dip in coherence at $\tau=\pi/\omega_L\approx235$~ns in the zero-detuning case, and two dips with a non-zero detuning. The results for a representative NV centre are shown in Fig.~\ref{FigCPMG}a, where the $\pi$-pulse duration was fixed to $t_p=40$~ns and the detuning varied from $\Delta/2\pi=+3$~MHz to $-3$~MHz in steps of 0.5~MHz. The data indeed reveals two dips at times $\tau_1$ and $\tau_2$ for non-zero detunings, and reproduces the fact that the amplitude of the $\tau_2$ dip (corresponding to the $\omega_1$ peak in the filter function) is larger than for $\tau_1$ at small detunings.

A direct comparison of theory and experiment is shown in Fig.~\ref{FigCPMG}b, which plots $\tau_1$ and $\tau_2$ as a function of the detuning as extracted from Fig.~\ref{FigCPMG}a (blue data) and as calculated for this situation (red lines). While there is a good qualitative agreement including the roll-over of $\tau_2$ near $|\Delta/2\pi|=2.5$~MHz, the splitting observed experimentally is larger than predicted. We repeated these experiments on several NV centres and systematically observed this trend. We attribute this discrepancy to a combination of pulse errors yet to be identified. We checked theoretically that a flip-angle error (i.e. an error in the choice of $t_p$) and/or a more realistic pulse shape do not produce a larger splitting. In Fig.~\ref{FigCPMG}c, we plotted the dip positions for a constant detuning $\Delta/2\pi=+1.5$~MHz but varying pulse duration $t_p$ from 80 ns to 24 ns (the smallest pulse duration achievable with our set-up), confirming the prediction that the splitting increases with $t_p$, although the measured splitting is again larger than predicted. 

It can be noted that Fig.~\ref{FigCPMG}a displays a small asymmetry  about $\Delta = 0$ whereby at $+0.5$~MHz the dip is split but at $-0.5$~MHz the dip is not split. We attribute this to experimental error. In particular, as explained in Sec.~\ref{sec:meth exp}, the detuning was observed to vary during the acquisition by up to 200~kHz due to residual temperature fluctuations. In this case, it is possible that the detuning of $-0.5$~MHz (average of the detuning measured before and after the acquisition) was in fact close to zero during a large portion of the acquisition, resulting in a single dip.
Moreover, in Fig.~\ref{FigCPMG} (theory and experiment) the splitting of the coherence dip is not symmetric about the central position, as it is in Fig.~\ref{fig: CPMG}c. This mismatch arises simply from plotting the data in different domains - frequency and (temporal) pulse spacing. Whilst the filter function splitting is symmetric about the central resonance position the switch to a scan over pulse spacing breaks this symmetry. Additionally, the splitting of the filter function is itself dependent on the pulse spacing which further contributes to the aberration. The theory dip positions presented in Fig.~\ref{FigCPMG}b represent a direct map from the filter function presented in Fig.~\ref{fig: CPMG}c.

The splitting of the CPMG filter function as a function of detuning explains why the CPMG protocol is rarely used for the spectroscopy of weak narrow-band signals~\cite{staudacher2013nuclear} (i.e. requiring a large number of $\pi$ pulses resulting in a narrow-band filter function), since the shape of the main resonance is directly affected by inhomogeneous broadening or drifts (resulting in a broadening of the NMR dip), and the splitting induced by a static detuning can even be mistaken for two NMR dips from two different nuclear spin species. This effect may also be relevant to broadband noise spectroscopy where CPMG is sometimes used~\cite{Myers2014,Romach2015} and $\tau$ is scanned over a large range (limited by $t_p$ for the highest frequencies), implying that the shape of the filter function is not constant across the scan and possibly resulting in spectral distortions.

\section{Effects of detuning in XY8 based NMR}
\label{sec: XY8}

We now investigate the effect of detuning on the XY8 sequence. Figure~\ref{FigXY8}a shows the filter function $|\tilde{\textbf{f}}(\omega)|^2$ for a constant pulse duration $t_p=40$~ns and different detunings. Unlike for the CPMG sequence, here the filter function exhibits a single resonance at $\omega_\text{DD}=1/2\tau$ for detunings up to 2~MHz, with only a small splitting appearing at $\Delta/2\pi=2.5$~MHz. This is due to the robust design of the XY8 sequence as discussed in Appendix~\ref{app: pulse prop} where the modulation functions are presented. The amplitude of the single peak decreases when the detuning increases, with a 6\% and 30\% reduction at $\Delta/2\pi=1.5$ and 2.0~MHz, respectively. 

We tested these detuning effects on the same NV as in Fig.~\ref{FigCPMG}a, with XY8-NMR spectra shown in Figs.~\ref{FigXY8}b (positive detunings) and \ref{FigXY8}c (negative detunings). There is indeed a clear reduction in the amplitude of the dip as the detuning is increased. Moreover, the presence of a detuning reduces the baseline coherence and add modulations that interfere with the NMR dip especially at large detunings (in particular, the NMR dip is no longer resolved at $\Delta/2\pi=-2.5$~MHz). 
These background coherence modulations arise from the detuning error effect on the qubit evolution even in the absence of dynamic noise (i.e. when $\Prop_p(t_\text{total})\ket{\psi(0)} \neq \ket{\psi(0)}$), and as such are not captured by Eq.~(\ref{eq: L classical}).
Experimentally, this is best seen in full-range decoherence curves (Fig.~\ref{FigXY8}d), where the presence of a detuning is seen to shorten the decoherence time $T_2$ and add large modulations that could be mistaken for NMR resonances. In contrast, the CPMG sequence is overall more robust to the detuning error, as can be seen by the $\Delta/2\pi=+3$~MHz curve showing a similar $T_2$ as the zero-detuning XY8 case, with no significant modulations. This is consistent with previous works that found the CPMG sequence to perform better than XY8 in protecting a qubit with an initial state parallel to the axis of the CPMG $\pi$-pulses~\cite{Ryan2010,Wang2012,Shim2012,Ali2013,Farfurnik2015}.

To illustrate how these detuning effects can affect the interpretation of NMR data, we used the model of Ref.~\cite{pham2016nmr} (which assumes no detuning) to fit the experimental data and extract the dephasing time, $T_{2n}^*$, of the ensemble of nuclear spins producing the NMR signal (related to the width of the NMR dip), and the depth of the NV centre, $d_{\rm NV}$ (related to the amplitude for a given width). These parameters are plotted against the detuning $\Delta$ in Fig.~\ref{FigXY8}e, showing variations significantly larger than the uncertainty (from the fit). For instance, this model estimates $T_{2n}^*=24\pm7~\mu$s and $d_{\rm NV}=11.8\pm0.3$~nm from the $\Delta/2\pi=+2$~MHz data, against $T_{2n}^*=12\pm1~\mu$s and $d_{\rm NV}=8.9\pm0.1$~nm at $\Delta=0$. We stress that such detunings are sometimes unavoidable, due to inhomogeneous broadening (a 2~MHz ODMR linewidth is not uncommon in dense layers of near-surface NV centres~\cite{Tetienne2018}) or hyperfine shifts, and as such this motivates the inclusion of detuning in the analysis of NMR data, or the development of pulse sequences that are less sensitive to detuning~\cite{casanova2015robust,Frey2017,Ziem2018}.

\section{Conclusion} \label{sec:conclusion}

In this work, we studied the effect of detuning errors in the context of NMR spectroscopy based on dynamical decoupling sequences applied to a qubit such as the NV centre in diamond. We found that the combination of non-zero detuning and finite pulse-duration gives rise to a splitting of the main resonance in the filter function for the CPMG sequence, and a modulation of the resonance amplitude for the XY8 sequence. These findings show that detuning errors, which are often unavoidable in experiments, must not be neglected when extracting quantitative information from NMR data, such as the number of spins in the sample or the depth of the NV centre. While in this paper we focused on two of the simplest dynamical decoupling sequences, CPMG and XY8, it would be interesting to investigate the effect of detunings on quantitative NMR sensing based on more advanced protocols specifically designed to be more robust against pulse errors~\cite{casanova2015robust,Frey2017,Ziem2018}. Another direction of interest is the study of different combinations of pulse errors in the NMR context, such as finite pulse-durations combined with a flip-angle error (i.e. pulses that are not exactly $\pi$). Such studies will shed light into the optimal conditions to perform accurate, quantitative NMR spectroscopy, and may also unveil new ways to extract useful information from the qubit coherence data.

\section*{Acknowledgements}

We acknowledge useful discussions with N. Bar-Gill and N. de Leon. This work was supported by the Australian Research Council (ARC) through grants CE110001027, FL130100119, DE170100129. J.E.L. is supported by a UK EPSRC DTA studentship. D.A.B. is supported by an Australian Government Research Training Program Scholarship. T.T. acknowledges the support of Grants-in-Aid for Scientific Research (Grant Nos. 15H03980, 26220903, and 16H06326), the "Nanotechnology Platform Project" of MEXT, Japan, and CREST (Grant No. JPMJCR1773) of JST, Japan.
 
\appendix

\section{Semi-classical model for finite-duration-pulse control}
\label{app: classical}

Here we detail the transformation to the toggling frame and the derivation of the analytic expression for the sensor qubit coherence response, Eq.~\eqref{eq: L classical}.

A classical signal felt by the qubit (here the NV electronic spin), can be modelled as a time-dependent magnetic field, $B_z(t)$. The microwave control is applied resonantly (plus some detuning) with one of the NV transitions to isolate a sensor qubit. In the frame rotating with the microwave drive frequency (and neglecting counter rotating terms) the Hamiltonian is given by  
\begin{equation}
\Ham_0(t) = -\gamma_\text{e}B_z(t)\Sz + \Ham_p(t),
\end{equation}
where $\gamma_\text{e} = -28\text{ GHz/T}\times 2\pi$ is the NV electronic gyromagnetic ratio and
\begin{equation}
\Ham_p(t) = \Delta \Sz + \sum_{m = 1}^N \Omega(t - t_m) \hat{S}_{\phi_m}
\end{equation}
is the microwave pulse Hamiltonian which describes a series of microwave pulses at the positions $t_m$ with phases $\phi_m$ and with shape $\Omega(t)$. The pulse shape $\Omega(t')$ is defined on the interval $t' \in [ -t_p/2, t_p/2 ]$ and we require $\int_{-t_p/2}^{t_p/2}\Omega(t)dt = \pi$ for a complete $\pi$-rotation. ($\hat{S}_{\phi_m} = (\exp(i\phi_m)\ketbra{d}{u} + \exp(-i\phi_m)\ketbra{u}{d})/2$ where $\ket{u,d}$ are the \textit{up} and \textit{down} states of the sensor qubit.) It is common to model the pulse shape with a delta-spike so that $t_p = 0$, however, here we assume some finite (non-zero) pulse duration and model the pulses as square (or top-hat) so that $\Omega(t) \equiv \Omega = \pi/t_p$. The Hamiltonian is presented here in the frame rotating with the microwave drive frequency and after making the rotating wave approximation to neglect the counter-rotating terms. The microwave drive is detuned from the NV resonance by $\Delta$. 

In the toggling frame (the frame rotating under the pulse propagator) the Hamiltonian can be written as
\begin{equation}
\Ham(t) = -\frac{\gamma_\text{e}}{2}B_z(t)\sum_i f_i(t) \sigi,
\end{equation}
which is presented in the main text. Here, $\sum_{i = x,y,z} f_i(t)\sigi = \Prop_p^\dagger(t) \sigz \Prop_p(t)$ with $\Prop_p(t) = \hat{\mathcal{T}}\exp(-i\int_0^t \Ham_p(s)ds)$ the pulse propagator. The transformation to this frame and the generalised modulation functions, $f_i(t)$, have been discussed previously \cite{lang2017enhanced} but here we include a detuning in the pulse Hamiltonian to obtain the modulation functions seen in Fig.~\ref{fig: CPMG}a. We use the Magnus expansion \cite{magnus1954exponential, albrecht2015filter, ma2016angstrom} to obtain an approximate evolution operator
\begin{align}
\Prop(t_\text{total}) &\approx \exp(-\frac{\gamma_\text{e}}{2}\sum_i \int_0^{t_\text{total}} f_i(t)B_z(t) dt \sigi)  \\
&\equiv \exp(-i\frac{1}{2}\sum_i \beta_i \sigi),  \label{eq: Uapprox}
\end{align}
where $\beta_i = -\gamma_\text{e} \int_0^{t_\text{total}} f_i(t)B_z(t) dt$. 

The coherence response is calculated via $\mathcal{L}(t_\text{total}) = \braket{\psi(t)|\sigx|\psi(t)}$ where we use Eq.~\eqref{eq: Uapprox} to evolve the initial sensor superposition state, $\ket{\psi(0)} = (\ket{u} + \ket{d})/\sqrt{2}$. (The true state evolution is determined by the combined evolution $\Prop(t_\text{total})\Prop_p(t_\text{total})$ as we are working in the toggling frame. However, we assume that the detuning is not so large so that $\Prop_p(t_\text{total})\ket{\psi(0)} \simeq \ket{\psi(0)}$ as discussed in Section~\ref{sec:meth theory}. The failure of this approximation at large detunings may account for the collapse of the coherence background seen in Figs.~\ref{FigXY8}b and \ref{FigXY8}c.) We find that 
\begin{align}
\mathcal{L}(t_\text{total}) 
&\simeq \braket{ 1 - 2\frac{\beta_y^2 + \beta_z^2}{|\pmb{\beta}|^2}\sin^2 \frac{1}{2}|\pmb{\beta}| }  \\
&\simeq \braket{ \cos |\pmb{\beta}| }
\end{align}
where in the last line we have assumed $\beta_x^2 \ll |\pmb{\beta}|^2$ and we define $|\pmb{\beta}|^2 = \beta_x^2 + \beta_y^2 + \beta_z^2$. 

We then follow the derivation in \cite{pham2016nmr} to find the expression presented in the main text,
\begin{equation}
\mathcal{L}(t_\text{total}) = 
\exp(-\frac{1}{2}\frac{\gamma_\text{e}^2}{2\pi}\int_{-\infty}^\infty S(\omega)|\tilde{\textbf{f}}(\omega)|^2 d\omega \ t_\text{total}^2),
\label{eq: L classical app}
\end{equation}
where $S(\omega)$ is the classical noise spectrum, $\tilde{\textbf{f}}(\omega) = (\tilde{f}_x(\omega), \tilde{f}_y(\omega), \tilde{f}_z(\omega))$ and we have defined the Fourier transform $\tilde{f}_i(w) = \frac{1}{t_\text{total}}\int_0^{t_\text{total}}f_i(t)\exp(-i\omega t)dt$. For $t_p = 0$ Eq.~\eqref{eq: L classical app} reduces exactly to the expression given in \cite{pham2016nmr}. 

\begin{figure}[t!]
\begin{center}
\includegraphics[width=0.99\columnwidth]{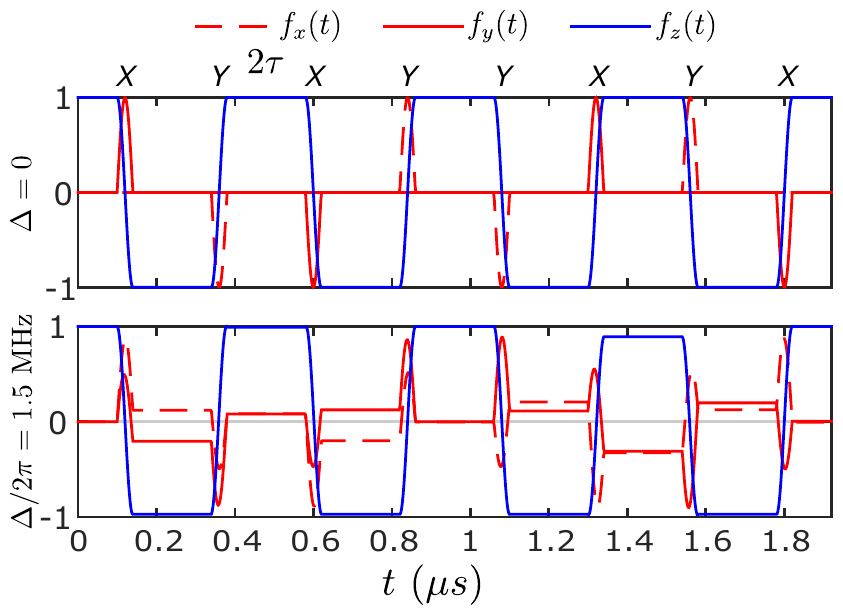}
\caption{Modulation functions $(f_x,f_y,f_z)$ in the time domain for the XY8 sequence with finite duration pulses for zero and non-zero detuning error. The simulation here has pulse spacing $\tau = 240$ ns, pulse width $t_p = 40$ ns and detunings $\Delta = 0$ (top graph) and $\Delta/2\pi = 1.5$ MHz (bottom) - matching the parameters as for the CPMG sequence in Fig.~\ref{fig: CPMG}a.}
\label{fig: XY8 modfuncs}
\end{center}
\end{figure}

\section{The XY8 sequence}
\label{app: pulse prop}

The XY family of DD sequences were designed specifically to protect an arbitrary initial state \cite{maudsley1986modified, gullion1990new}. By design the XY8 sequence is more robust than CPMG and the error only accumulates at second order \cite{wang2012comparison}. However, this error generates a rotation about the $(\hat{\pmb{x}} + \hat{\pmb{y}})/\sqrt{2}$-axis which is not parallel with the initial state $(\ket{u} + \ket{d})/\sqrt{2}$. We attribute the large modulations of the XY8 background coherence in Fig.~\ref{FigXY8} to this fact. If enough pulses are applied the small error can be compounded, whereas for CPMG the first order $x$-rotation actually works to minimise the modulations created by higher order terms as it stabilises the rotation axis around the $x$-axis. This is consistent with previous studies showing that CPMG can outperform XY8 at protecting states aligned with the $x$-axis \cite{Ryan2010,Wang2012,Shim2012,Ali2013,Farfurnik2015}. Figure~\ref{fig: XY8 modfuncs} shows the XY8 modulation functions for the same parameters as in Fig.~\ref{fig: CPMG}a. Whilst the detuning error alters the modulation functions slightly the sequence corrects these alterations by the end of the sequence, at $t = 8\tau$. This explains why the filter function splitting is not seen for the XY8 sequence (until larger detuning strengths).

\bibliographystyle{unsrt}
\bibliography{bib}

\end{document}